\documentclass[11pt,a4paper]{article}
\usepackage{jcappub}
\usepackage{latexsym,hyperref}
\usepackage{jcappub}

\newcommand{\be}{\begin{equation}}
\newcommand{\ee}{\end{equation}}
\newcommand{\ba}{\begin{eqnarray}}
\newcommand{\ea}{\end{eqnarray}}
\newcommand{\beq}{\begin{equation}}
\newcommand{\eeq}{\end{equation}}
\newcommand{\bea}{\begin{eqnarray}}
\newcommand{\eea}{\end{eqnarray}}
\newcommand{\bi}{\begin{itemize}}
\newcommand{\ei}{\end{itemize}}
\newcommand{\bfi}{\begin{figure}[!t]
\epsfxsize=9cm
\epsffile}
\newcommand{\efi}{\end{figure}}
\newcommand{\efig}{\end{figure*}}
\newcommand{\no}{\nonumber}
\newcommand{\non}{\nonumber}

\newcommand{\bfs}{{\bf s}}
\newcommand{\bfx}{{\bf x}}
\newcommand{\bfk}{{\bf k}}
\newcommand{\bfr}{{\bf r}}
\newcommand{\bfv}{{\bf v}}

\newcommand{\bfn}{{\bf n}}

\title{Quantification of the multi-streaming effect in Redshift Space Distortion}
\author[a]{Yi Zheng,}
\author[b,c]{Pengjie Zhang}
\author[a,d]{and Minji Oh}

\affiliation[a]{Korea Astronomy and Space Science Institute, 776, Daedeokdae-ro, Yuseong-gu, Daejeon 34055, Republic of Korea}
\affiliation[b]{Center for Astronomy and Astrophysics, Department of Physics and Astronomy, Shanghai Jiao Tong University, Shanghai, 200240, China}
\affiliation[c]{IFSA Collaborative Innovation Center, Shanghai Jiao Tong University, Shanghai, 200240, China}
\affiliation[d]{University of Science and Technology, Daejeon 34113, Korea}

\emailAdd{yizheng@kasi.re.kr}
\emailAdd{zhangpj@sjtu.edu.cn}

\abstract{Both multi-streaming (random motion) and bulk motion cause the Finger-of-God (FoG) effect in redshift space distortion (RSD).  We apply a direct measurement of the multi-streaming effect in RSD from simulations, proving that it induces an additional, non-negligible FoG damping to the redshift space density power spectrum. We show that, including the multi-streaming effect,  the RSD modelling is significantly improved. We also provide a theoretical explanation based on halo model for the measured effect, including a fitting formula with one to two free parameters.  The improved understanding of FoG helps break the $f\sigma_8-\sigma_v$ degeneracy in RSD cosmology, and has the potential of significantly improving cosmological constraints.} 

\begin{document}
\maketitle
\flushbottom


\section{Introduction}
\label{sec:intro}

Redshift space distortion (RSD) \cite{Jackson72,Sargent77,Peebles80,Kaiser87,Peacock94,Ballinger96} as a cosmological probe is invaluable. Physically speaking, it offers the possibility to statistically measure the peculiar velocity at cosmological distances without the problem of subtracting the Hubble flow. Therefore it becomes one of the major tools to probe dark matter, dark energy and gravity at cosmological scales. Observationally, RSD has been robustly measured for over a decade and will be measured to $1\%$ precision level over the coming decade(e.g. \cite{Peacock01,Tegmark02,Tegmark04,Samushia12,Guzzo08,Blake11b,Blake12,Marin16,Beutler16,Hector16,Satpathy16,PFS12,Euclid13,Zhao16,DESI16I}).

Accurate modelling of the RSD effect is crucial to gaining tighter cosmological constraints and making the best use of the fund spent on the future galaxy redshift surveys. However, the RSD theoretical modelling is highly challenging, in particular in the non-linear regime and at $1\%$ accuracy level \cite{Peebles80,Fisher95,Heavens98,White01,Seljak01,Kang02,Tinker06,Tinker07,
Scoccimarro04,Matsubara08a,Matsubara08b,Desjacques10,Taruya10,Taruya13,
Matsubara11,Okumura11a,Okumura11b,Sato11,Jennings11b,
Reid11,Seljak11,Okumura12,Okumura12b,Kwan12,Zhangrsd,Zheng13,Ishikawa14,White15,Jennings16,Bianchi15,Bianchi16,Simpson16}. There are three major obstacles to overcome: the non-linear mapping from real to redshift space, the failure of perturbation theory to describe the non-linear dark matter density and velocity evolutions at small scales, and the complexities in  the galaxy-halo relation and the resulting uncertainties in modelling galaxy density and velocity biases.  Each of them has many model uncertainties. When we combine them together into the RSD modelling, it results in an awkward situation. When the RSD model works badly, we can not easily tell which part should be mostly to blame. Even when the model works well, we are not sure if the model is physically correct, or just a coincidence where several errors happen to cancel each other.  

Therefore, we take an alternative approach in RSD study. We disentangle these complexities of RSD modelling into separate parts and try to improve the physical understanding of each part, in the context of RSD. On the galaxy and halo bias part, we have measured the volume-weighted halo velocity bias in \cite{Zheng14b}. To correct the otherwise severe sampling artifact entangled in the volume-weighted velocity statistics measurement, we have developed a theoretical model of  the sampling actifact \cite{Zhang14}, tested it against N-body simulations \cite{Zheng14a} and then applied it to the halo velocity bias measurement \cite{Zheng14b}.  On the real space dark matter evolution part, we have measured,  through a set of N-body simulations, the non-linear evolution of the dark matter velocity field and its relation with the dark matter density field \cite{Zheng13}. The result of non-linear density-velocity relation has been used in BOSS data analysis \cite{Wang16,Zhigang16} and is proved to be robust in measuring the cosmological structure growth rate.  

On the real space-redshift space mapping, we have proposed a new velocity decomposition scheme which leads to a new model of mapping in \cite{Zhangrsd}. Part of the predictions of this model have been tested against dark matter RSD in N-body simulations  \cite{Zheng13} and the rest will be tested in a companion paper. Meanwhile,  one of the authors has tested the TNS-based \cite{Taruya10} dark matter clustering mapping formula from real to redshift space in \cite{Zheng16a}. The halo density bias and the halo clustering mapping formula of this model will be tested in \cite{Zheng16c}. However, these investigations so far overlook an important physical process of non-linear structure formation, namely the shell crossing and the resultant multi-streaming effect. The current paper quantifies its influence on the redshift space clustering using N-body simulations.

Non-linear evolution of the density field induces multi-streaming effect at small scales after shell crossing. In principle, the multi-streaming effect in RSD modelling can be dealt with by the distribution function approach (hereafter DFA) \cite{Seljak11,Okumura12}.  However so far it is still difficult to account for the multi-streaming effect in the language of perturbation theory and DFA. On the other hand,  when we test the dark matter mapping formula from simulations, we try to sample the volume-weighted velocity field.\footnote{If we sample the mass-weighted velocity fields from simulations, the multi-streaming effect is already contained in the measurement. But the halo momentum is not conservative, so it would be difficult to calculate the halo momentum field from theory, but see \cite{Sugiyama16}.}  This sampling by definition produces a single-valued field on the grid and therefore ignores the multi-streaming at scales of the grid size and below. As shown later, this will lead to an underestimation of the Finger-of-God (hereafter FoG) effect.  Roughly speaking, the velocity disperion responsible for the FoG has two sources, which often cause confusions. The original ansatz of FoG targeted at the multi-streaming effect. Therefore the velocity dispersion $\sigma_v$ inferred from the FoG is often interpreted as the random motions inside halos. However, later studies (e.g. \cite{Scoccimarro04,Zhangrsd}) showed that the bulk flow also causes significant FoG effect. In this case, the corresponding $\sigma_v$ is completely determined by the velocity power spectrum and therefore can provide useful constraint of cosmology. The actual FoG is the combination of the two. Ignoring the multi-streaming might mislead our judgement about the accuracy of RSD model. It also misleads the interpretation of $\sigma_v$ and its cosmological constraint. 

Combining N-body simulations and the otherwise exact RSD modelling under the single-streaming approximation, we are able to quantify the multi-streaming effect in the context of RSD. In section~\ref{sec:measure}, we measure the multi-streaming induced damping of redshift space clustering from N-body simulations. In section~\ref{sec:reexam}, we show that it provides the missing ingredient in explaining a puzzle found in the RSD model test of \cite{Zheng16a}, and can therefore significantly improve the RSD model accuracy. In section~\ref{sec:halomodel}, we explain the finding of multi-streaming with the halo model, and provide a fitting formula with up to two free parameters.  We conclude and discuss in section~\ref{sec:con}. 
\section{Measurement from simulations}
\label{sec:measure}

In this section, we provide a direct measurement of the multi-streaming damping in dark matter RSD effect. The estimator is straightforward to define from the derivation of the RSD mapping formula from real space to redshift space.

\subsection{Estimator definition}
\label{subsec:theory}

Under the plane parallel approximation, for distant tracer, the real space  position $\bfx$ is related to its redshift space position $\bfs$ as 
\beq
\label{eq:mapping}
\bfs=\bfx+\frac{v_z(\bfx)}{H(z)}\widehat{z}\,,
\eeq
in which we adopt $\widehat{z}$ axis as the line of sight, $v_z$ is the comoving velocity component along the line of sight which is far below the speed of light, and $H(z)$ is the Hubble parameter at redshift $z$.

Suppose there are $j=1,\cdots, N$ tracers (simulation particles/halos/galaxies) in a volume of $V$ with the averaged number density $\bar{n}=N/V$, the number overdensity in redshift space is 
\be
\delta^s({\bf s})=\frac{1}{\bar{n}}\sum_j \delta_{3D}\left[{\bf s}-({\bf x}_j+\frac{v_{z,j}}{H}\hat{z}_j)\right]-1\,.
\ee
Its Fourier counterpart is 
\ba
\label{eqn:deltas1}
\delta^s({\bf k})&=&\frac{1}{\bar{n}}\sum_j e^{ik_zv_{z,j}/H} e^{i{\bf k}\cdot{\bf x}_j}-(2\pi)^3\delta_{3D}({\bf k}) \no \\
&=& \frac{1}{\bar{n}}\sum_j e^{ik_zv_{z,j}/H} e^{i{\bf k}\cdot{\bf x}_j}\ \ {\rm when}\ {\bf k}\neq {\bf 0}\,.
\ea
Here $k_z\equiv k\mu$ is the $k$ mode component along the line of sight and $\mu$ is the cosine of the angle between $\bfk$ and $\widehat{z}$.
Since we have no interest in the case of ${\bf k}={\bf 0}$, hereafter we will adopt the last expression.  The redshift space power spectrum is then
\be
\langle \delta^s({\bf k}) \delta^s({\bf k}^{'})\rangle\equiv (2\pi)^3\delta_{3D}({\bf k}+{\bf k}^{'}) P^s_{\rm multi}({\bf k})\,.
\ee
The subsctript ``multi'' of the power spectrum emphasizes that it automatically takes the possible multi-streaming into account, in contrast to the later discussed $P^s_{\rm single}({\bf k})$, which assumes single-streaming. This power spectrum can be directly measured from the simulated/observed redshift space distribution of particles/galaxies. Or it can be calculated following eq.~(\ref{eqn:deltas1}). The two approaches are mathematically equivalent.

While $P^s_{\rm multi}({\bf k})$ is most straightforwardly measured from simulations/data, the theoretical RSD modelling usually utilizes the single-streaming approximation and considers the velocity field as a single valued field.\footnote{But see \cite{Colombi15,Taruya16} for pioneering studies of multi-streaming perturbation theory.} We denote the resultant power spectrum as $P^s_{\rm single}({\bf k})$.   The same approximation is adopted when we separately sample density and volume-weighted velocity fields in simulations. For example, all velocity assignment methods force a single-valued velocity field, yet could be discontinuous. When such approximation is adopted, the discrete sum $\sum_j/\bar{n}$ can be replaced by the integral $\int d^3{\bf x}(1+\delta({\bf x}))$ and we obtain the familiar form 
\beq
\label{eq:deltas}
\delta^s _{\rm single} (\bfk)=\int d^3\bfx\exp(i\bfk\cdot\bfx)[1+\delta(\bfx)]\exp\left(i\frac{k_zv_z}{H}\right) \,.
\eeq
 The density power spectrum in redshift space is then
\bea
\label{eq:psred}
P^s_{\rm single}&=&\int d^3\bfr\exp(i\bfk\cdot\bfr)\left\langle (1+\delta_1)e^{i\frac{k_zv_{1z}}{H}}(1+\delta_2)e^{-i\frac{k_zv_{2z}}{H}}\right\rangle \,. \non
\eea
Here $\delta_i\equiv\delta(\bfx_i)$, $v_{i,z}\equiv v_z(\bfx_i)$, and $\bfr\equiv\bfx_1-\bfx_2$.  Notice that the velocity field is  single-valued and therefore multi-streaming is indeed neglected in the above expression.


Multi-streaming means that particles at the same position can have different velocities. It can be non-negligible in the non-linear regime and even be significant in virialized halos. 
To deal with multi-streaming, $\sum_j/\bar{n}$ in eq.~(\ref{eqn:deltas1}) should be replaced by $\int d^3{\bf x}d^3{\bf v} f({\bf x},{\bf v})/\bar{n}$ where $f({\bf x},{\bf v})$ is the phase space distribution of particles. The particle velocity at position ${\bf x}$ then has two components, the bulk velocity $\bar{\bf v}\equiv \int {\bf v}f({\bf x},{\bf v}) d^3{\bf v}/\int f({\bf x},{\bf v}) d^3{\bf v}$ and the random velocity $\delta {\bf v}\equiv {\bf v}-\bar{\bf v}$. Eq.~(\ref{eqn:deltas1}) is then
\ba
\label{eqn:deltas2}
\delta^s({\bf k})&=&\frac{1}{\bar{n}}\int d^3{\bf x} e^{i{\bf k}\cdot{\bf x}} e^{ik_z\bar{v}_z({\bf x})/H}d^3{\bf v}f({\bf x},{\bf v})e^{ik_z\delta v_z/H} \no \\
&=&\int d^3{\bf x} e^{i{\bf k}\cdot{\bf x}} e^{ik_z\bar{v}_z({\bf x})/H} [1+\delta({\bf x})] \left\langle e^{ik_z\delta v_z/H}\right\rangle_{{\bf v}}\,.
\ea
The above expression uses the definition  $\int d^3{\bf v}f({\bf x},{\bf v})=\bar{n}(1+\delta({\bf x}))$. The average $\langle A\rangle_{\bf v}\equiv \int d^3{\bf v}f({\bf x},{\bf v}) A({\bf x},{\bf v})/\int d^3{\bf v}f({\bf x},{\bf v})$ is over the velocity distribution at fixed ${\bf x}$. Eq.~(\ref{eqn:deltas2}) reduces to eq.~(\ref{eq:deltas}) in the limit of single-streaming ($|\delta \bfv|\rightarrow 0$ and therefore $e^{ik_z\delta v_z/H}\rightarrow 1$). Taking into account of  multi-streaming and applying the cumulant expansion theorem, we have
\ba
\label{eqn:deltamulti}
\delta^s _{\rm multi} ({\bf k})&=&\int d^3{\bf x} e^{i{\bf k}\cdot{\bf x}} e^{ik_z\bar{v}_z({\bf x})/H} [1+\delta({\bf x})]\no \\
&&\times  \exp\left[-\frac{k^2_z\langle \delta v_z^2\rangle_{\bf v}({\bf x})}{2H^2}+\cdots\right]\,.
\ea

As expected, multi-streaming mixes phases of the density field and therefore damps the redshift space clustering. It results in 
\beq
\label{eq:Dmulti}
D_{\rm multi}({\bf k})\equiv \frac{{P}^s_{\rm multi}({\bf k})}{{P}^s_{\rm single}({\bf k})} <1\,.
\eeq
An immediate task is then to quantify $D_{\rm multi}({\bf k})$, the impact of multi-streaming on the redshift space power spectrum.

One may wonder whether there exists a single-valued velocity assignment method which can take multi-streaming into account and make $D_{\rm multi}=1$.  Suppose that such method exists and the assigned value is ${\bf v}^{\rm assign}$, then it must satisfy  $\int d^3{\bf v} f({\bf x},{\bf v})\exp(ik_zv_z/H)=\exp(ik_zv^{\rm assign}_z/H) \int d^3{\bf v} f({\bf x},{\bf v})$. The r.h.s.  expression has an unique dependence on $k_z$.  However, when multi-streaming exists, there are other distinctive $k_z$ dependences (eq.~(\ref{eqn:deltamulti})). Therefore no ${\bf v}^{\rm assign}$ can satisfy the above condition for all $k_z$. It is then clear that any single-valued velocity assignment method miscaptures multi-streaming to some extent and leads to $D_{\rm multi}\neq 1$. 


\begin{table}[t]
\centering
\scriptsize
\begin{tabular}{@{}lll}
\hline\hline
parameter & physical meaning & value \\
\hline
$\Omega_m$  & present fractional matter density & $0.3132$ \\
$\Omega_{\Lambda}$ & $1-\Omega_m$ & $0.6868$ \\
$\Omega_b$ & present fractional baryon density & $0.049$\\
$h$ & $H_0/(100$~km~s$^{-1}$Mpc$^{-1})$ & $0.6731$ \\
$n_s$ & primordial power spectral index & $0.9655$ \\
$\sigma_{8}$ & r.m.s. linear density fluctuation & $0.829$ \\
\hline
$L_{\rm box}$ & simulation box size & 1890~$h^{-1}$Mpc\\
$N_{\rm p}$ & simulation particle number & $1024^3$\\
$m_{\rm p}$ & simulation particle mass & $5.46\times 10^{11}h^{-1}M_{\odot}$\\
\hline
$N_{\rm snap}$ & number of output snapshots & $13$ \\
$z_{\rm ini}$ & redshift when simulation starts & $49.0$ \\
$z_{\rm final}$ & redshift when simulation finishes & $0.0$ \\
\hline
\end{tabular}
\caption{The parameters and technical specifications of the N-body simulations for this work.}
\label{tab:simulation}
\end{table}

\subsection{$D_{\rm multi}$ measurement}
\label{subsec: measure}

We use the same set of simulations introduced in a previous paper \cite{Zheng16a} by one of the authors to measure $D_{\rm multi}$. These are 100 N-body dark matter simulations run by GADGET2 \cite{Springel05}, with box-size $L_{\rm box}=1.89\,h^{-1}$Gpc and $N_{\rm p}=1024^3$ particles. The box-size is chosen to mimic the survey volume that DESI will observe between $z=0.8$ and $z=1.0$ \cite{DESIwhite}. All simulations are generated by the same LCDM cosmology with Gaussian initial condition and flat space. The cosmological parameters are identical to PLANK15 results \cite{PLANK2015}. The initial conditions are generated by the 2LPT code \cite{2LPT} at $z=49$. We mainly analyse 4 snapshots in this paper, namely $z=0.0$, 0.5, 0.9, and 1.5. The detailed simulation parameters are listed in table \ref{tab:simulation}.

We use $N_{\rm s}=30$ simulations for calculation and error estimation. On a regular grid with $512^3$ grid points, we sample the density field by the clouds-in-cell (CIC) method and the velocity field by the nearest particle (NP) method \cite{Zheng13}. We correct the shotnoise and window function effect in the density field sampling and conduct 
a convergence test which guarantees that the sampled velocity field is accurate within $\sim1\%$ at $k\lesssim0.2h$/Mpc. Detailed numerical artifact treatments are described in Appendix \ref{appsec:numerical}.

For $P^s_{\rm multi}$, we first shift the positions of particles along the line-of-sight according to eq.~(\ref{eq:mapping}), then we sample the density field $\rho(\bfx)$ on $512^3$ regular grid points. We calculate the overdensity field $\delta(\bfx)=\rho(\bfx)/\bar{\rho}-1$ and use the fast Fourier transform (FFT) to calculate its Fourier counterpart $\delta(\bfk)$. Finally we average $\delta(\bfk)\delta^\ast(\bfk)$ in the same $(k,\mu)$ bins to generate $P^s_{\rm multi}(k,\mu)$.

For $P^s_{\rm single}$, we sample the real space overdensity $\delta(\bfx)$ and velocity $\bfv(\bfx)$ fields on $512^3$ regular grid points separately, and then make calculations through the following formula,
\bea
P^s_{\rm single}&=&\int d^3\bfr\exp(i\bfk\cdot\bfr)\left\langle (1+\delta_1)\exp(i\frac{k_zv_{1z}}{H})\times(1+\delta_2)\exp(-i\frac{k_zv_{2z}}{H})\right\rangle \non \\ 
&=&\int d^3\bfr\exp(i\bfk\cdot\bfr)\left\langle \delta_1\exp\left(i\frac{k_zv_{1z}}{H}\right)\times\delta_2\exp\left(i\frac{k_zv_{2z}}{H}\right)\right\rangle \non \\
&&+2\int d^3\bfr\exp(i\bfk\cdot\bfr)\left\langle \delta_1\exp\left(i\frac{k_zv_{1z}}{H}\right)\times \exp\left(i\frac{k_zv_{2z}}{H}\right)\right\rangle \non \\
&&+\int d^3\bfr\exp(i\bfk\cdot\bfr)\left\langle \exp\left(i\frac{k_zv_{1z}}{H}\right)\times\exp\left(i\frac{k_zv_{2z}}{H}\right)\right\rangle \non \\
&\equiv &P_1+P_2+P_3\,,
\label{eq:p_single}
\eea
where

\bea
\label{eq:separate_p}
P_1&\equiv&\int d^3\bfr\exp(i\bfk\cdot\bfr)\left\langle \delta_1\exp\left(i\frac{k_zv_{1z}}{H}\right)\times\delta_2\exp\left(i\frac{k_zv_{2z}}{H}\right)\right\rangle\,,\\
P_2&\equiv&2\int d^3\bfr\exp(i\bfk\cdot\bfr)\left\langle \delta_1\exp\left(i\frac{k_zv_{1z}}{H}\right)\times \exp\left(i\frac{k_zv_{2z}}{H}\right)\right\rangle\,, \non\\
P_3&\equiv&\int d^3\bfr\exp(i\bfk\cdot\bfr)\left\langle \exp\left(i\frac{k_zv_{1z}}{H}\right)\times\exp\left(i\frac{k_zv_{2z}}{H}\right)\right\rangle\,.\non
\eea
We separate the measurement into three parts, since different part has different numerical artifact corrections, which are detailed in Appendix \ref{appsec:numerical}. For each part, we first construct the fields $[\delta\exp(ik_zv_z/H)](\bfx)$ and/or $[\exp(ik_zv_z/H)](\bfx)$ from separately sampled $\delta(\bfx)$ and/or $\bfv(\bfx)$ fields, then calculate $P_{1,2,3}$ in the same way as we do for calculating the auto- or cross-power spectrum.

\begin{figure}
\centering
\includegraphics[width=0.49\columnwidth]{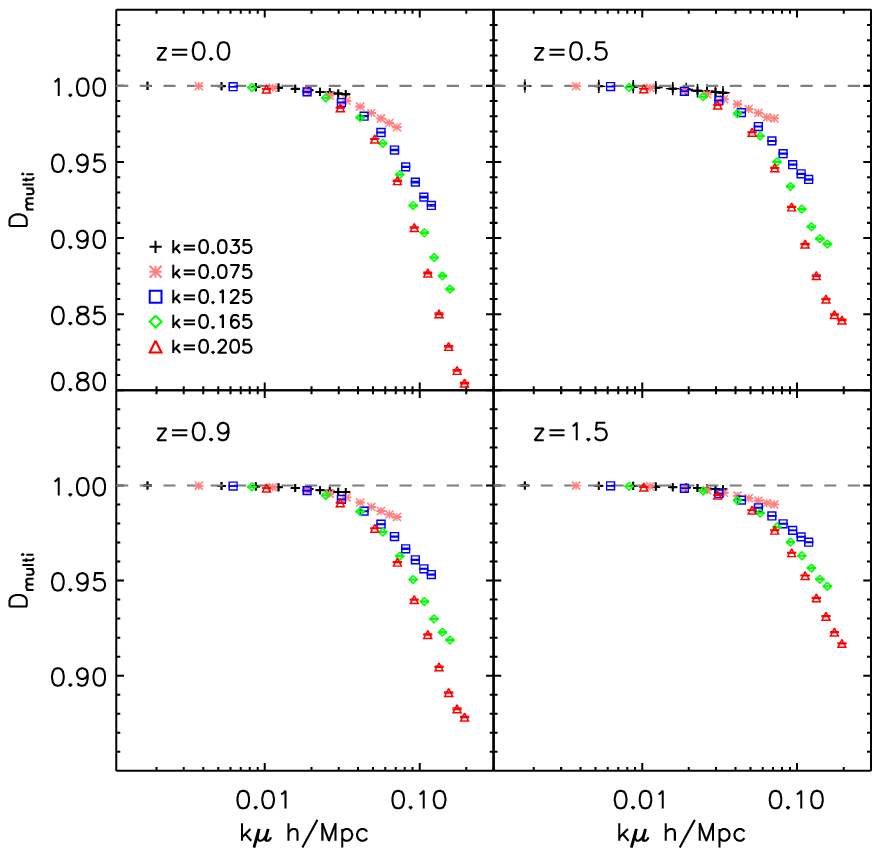}
\hfill
\includegraphics[width=0.49\columnwidth]{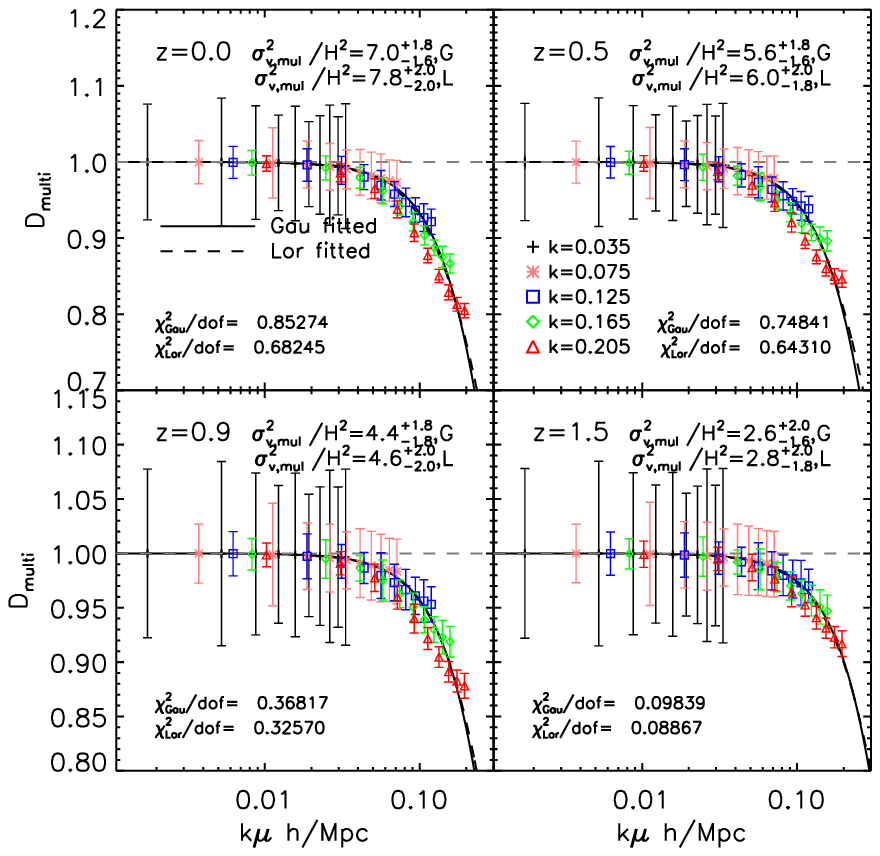}
\caption{The measurement of $D_{\rm multi}\equiv{P}^s_{\rm multi}/{P}^s_{\rm single}$ from simulations. The data points are the mean values of $D_{\rm multi}$ averaged over 30 simulations. {\em Left panel:} The error bars represent the standard errors of the mean of $D_{\rm multi}$ from 30 simulations. {\em Right panel:} The error bars come from the standard errors of the measured $P^s_{\rm multi}$ from 30 simulations. Solid/dashed lines are the Gaussian/Lorentzian fittings of the measured $D_{\rm multi}$. Notice that the $y$ ranges of left and right panels are different.}
\label{fig:d_measure}
\end{figure}

We take the ratio of $P^s_{\rm multi}$ and $P^s_{\rm single}$ to obtain $D_{\rm multi}$. The results are shown in figure~\ref{fig:d_measure}. The data points are the mean values of $D_{\rm multi}$ averaged over 30 simulations. We present two kinds of error bars in figure~\ref{fig:d_measure}. On the left panel,  the error bars represent the standard errors of the mean of $D_{\rm multi}$ measured from 30 simulations.\footnote{Denoting the standard error of measured $D_{\rm multi}$ across $N_{\rm s}$ simulations as $\sigma_{D,\rm multi}$, the standard error of the mean is $\sigma_{D,\rm multi}/\sqrt{N_{\rm s}}$.} This is to show the accurate measurement of $D_{\rm multi}$  for unveiling its real properties. The error bars are very small since both the numerator and denominator are measured from the same realization so the cosmic variance is cancelled out. On the right panel, the error bars come from the standard errors of $P^s_{\rm multi}$ measured from 30 simulations.\footnote{Denoting the standard error of $P^s_{\rm multi}$ across $N_{\rm s}$ simulations as $\sigma_{P,\rm multi}$, and the averaged $P^s_{\rm single}$ as $\langle P^s_{\rm single}\rangle$, the error bars on the right panel of figure~\ref{fig:d_measure} are $\sigma_{P,\rm multi}/\langle P^s_{\rm single}\rangle$.} This is for showing how well this damping effect could be modelled by a Gaussian or Lorentzian function considering the statistical errors of a survey with the same volume as our simulation.

(i) The damping amplitudes of the results clearly show that the multi-streaming effect causes an additional, non-negligible FoG effect. For example, $D_{\rm multi}\simeq 0.9$ at $k=0.1h$/Mpc of $z=0.0$ and $k=0.2h/$Mpc of $z=0.9$, presenting a $10\%$ damping to the redshift space clustering.  This damping effect deepens at lower redshifts, when the dark matter velocity dispersions inside halos increase due to the higher non-linearity. 

(ii) On the left panel, this damping function clearly deviates from the pure $k\mu$ dependence of widely adopted FoG function. The deviations are larger than percent level at $k\gtrsim 0.1h/$Mpc. This extra $k$ and/or $\mu$ dependence will affect the interpretation of the RSD model.\footnote{Basically the extra $k$ and/or $\mu$ dependence of $D_{\rm multi}$ goes into the FoG term and make $D^{\rm FoG}$ deviating from a pure function of $k\mu$. In figure 6 of \cite{Zheng16a}, the combination of higher order terms $A+B+T$ works better than $A+B+T+F$. This could be explained by that the $k$ and $\mu$ dependence of $F$ term (by accident) cancels the extra $k$ and/or $\mu$ dependence of $D_{\rm multi}$.} 

(iii) On the right panel, the solid/dashed lines are Gaussian/Lorentzian fittings of the measured $D_{\rm multi}$,
\bea
D_{\rm multi}=\left\{
\begin{array}{ll}
e^{-k_z^2\sigma_{v,\rm multi}^2/H^2(z)}& {\rm Gaussian \,, }  \\
\frac{1}{1+k_z^2\sigma^2_{v,\rm multi}/H^2(z)} & {\rm Lorentzian\,.}  \\
\end{array}\right. 
\eea
Then the damping effect effectively corresponds to a multi-streaming velocity dispersion, $\sigma_{v,\rm multi}$, whose fitted values with error bars are presented on the top right sides. We fit the data from $k=0.035h$/Mpc to $k=0.205h$/Mpc. The fitted $\sigma_{v,\rm multi}^2$ increases towards lower redshift, due to larger multi-streaming damping effect. Both fitted $\chi^2$ per degree of freedom are $\lesssim 1$, indicating that the Gaussian and Lorentzian functions are both good choices for describing the damping effect, considering our simulation volume.\footnote{The Lorentzian function has a slightly better fit to the data than the Gaussian function. In the language of halo model, this is due to the fact that all halos with different masses contribute to the multi-streaming damping \cite{White01,Seljak01} and their contributions should be integrated together to evaluate $D_{\rm multi}$ (eq.~(\ref{eq:pddred})). For simplicity, we will use the Gaussian function to describe the multi-streaming damping in this paper.} The theoretical description of multi-streaming damping by halo model will be discussed in section~\ref{sec:halomodel}.

\section{Improving the RSD modelling by including multi-streaming}
\label{sec:reexam}

\begin{figure}
\centering
\includegraphics[width=.65\textwidth]{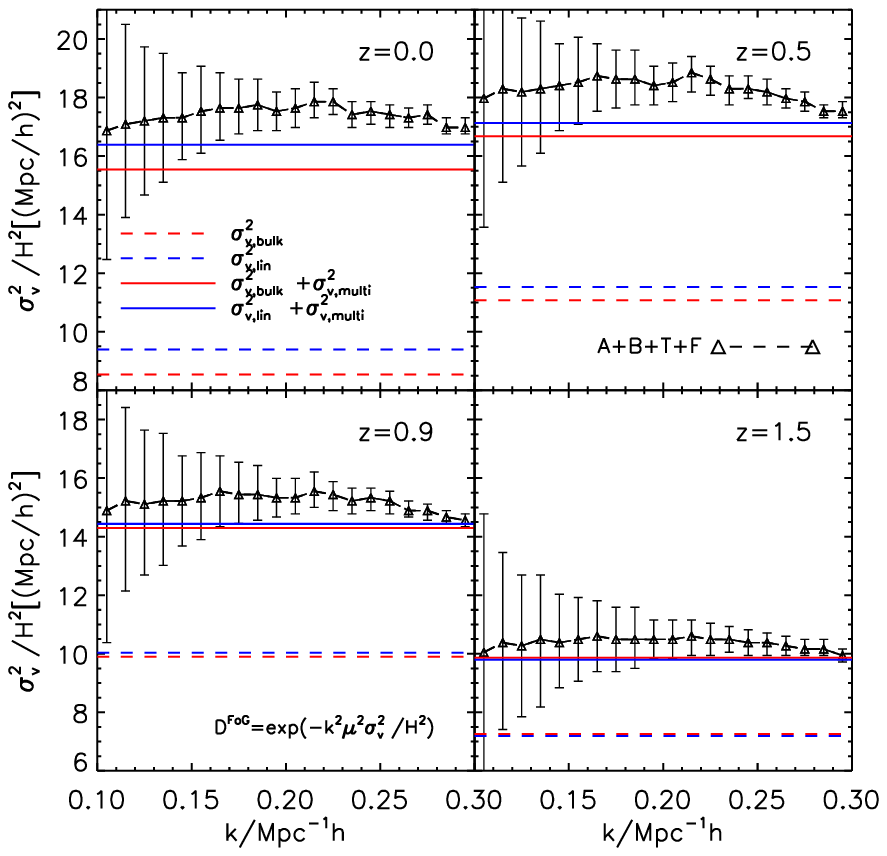}
\label{fig:sgmfit}
\caption{Fitted $\sigma_v^2/H^2(k)$  for the ``extended'' TNS model studied in \cite{Zheng16a}. For all terms needed in this fitting, the density field is sampled by CIC method, and the velocity field is sampled by NP method. The errors of the fitted $\sigma_{v,\rm multi}^2/H^2$ are shown in figure~\ref{fig:sgmv2}. We do not include them in this figure for the conciseness of the plot.}
\end{figure}
The redshift space power spectrum consists of a part of clustering enhancement (generalized Kaiser effect) and a part of clustering damping (FoG). Schematically, it can be written as (e.g. \cite{Zhangrsd})
\ba
P^s(k,u)={\rm Enhancement}(k,\mu)\times D^{\rm FoG}(k\mu)\,.
\ea
Robust measurement of FoG relies on robust modelling of the enhancement term, which contains various corrections to the Kaiser formula $(1+f\mu^2)P_{\rm lin}$ \cite{Kaiser87}. Notice that the damping term $D^{\rm FoG}$ only depends on the combination $k\mu$, instead of both $k$ and $\mu$. Only if we model the enhancement term correctly can the inferred $D^{\rm FoG}$ satisfy the above constraint. This provides a judgement on whether the inferred FoG is robust. \cite{Taruya10} proved that, expanding the mapping formula to 2nd order in $k\mu$ and only retaining $A$ and $B$ terms are enough to accurately predict the redshift space clustering on BAO scales. \cite{Zheng16a} investigated the TNS model \cite{Taruya10}  and found that,  including the complete terms to the 2nd order of Taylor expansion in terms of $k\mu$ ($A$, $B$, $F$, and $T$ terms) is crucial to accurately describe the redshift space clustering at $\lesssim0.2h$/Mpc. In this ``extended'' TNS model,
\bea
\label{eq:Pkred}
{\rm Enhancement}&=&[P_{\delta\delta}+2\mu^2P_{\delta\Theta}+\mu^4P_{\Theta\Theta}  \\
&&+A(k,\mu)+B(k,\mu)+T(k,\mu)+F(k,\mu)] \nonumber \,.
\eea
By approximating FoG as $D^{\rm FoG}=e^{-k^2\mu^2\sigma_v^2/H^2}$ and by fitting against the $\mu$ dependence of each $k$ bin,  the line-of-sight velocity dispersion $\sigma_v^2$ of each $k$ bin can be obtained (figure~\ref{fig:sgmfit}, or figure 8 of the original paper \cite{Zheng16a}).\footnote{The fitted data points in figure~\ref{fig:sgmfit} is slightly different from those of figure 8 of \cite{Zheng16a}, since here we use CIC method to sample the density field, which suffers less alias effect \cite{Jing05} than the NGP method adopted in \cite{Zheng16a}.}  In the fitting, the bin size is $\Delta k=0.01h$/Mpc. Figure~\ref{fig:sgmfit} shows that, the fitted $\sigma_v^2$ is indeed constant at $k\lesssim0.3h$/Mpc, which means that the inferred $D^{\rm FoG}$ indeed only depends on $k\mu$ and therefore the FoG measurement is robust.\footnote{In \cite{Zheng16a}, the extended TNS model is tested by full simulation calculation and proved to be accurate within $\lesssim 2\%$ at $k\lesssim0.2h$/Mpc in describing the full 2D redshift space power spectrum.}

However, the fitted $\sigma_v^2$ is significantly larger than $\sigma_{v,{\rm bulk}}^2$, obtained either from the linear theory (blue dashed lines), or from the volume-weighted velocity field on $512^3$ regular grid points by NP method in simulations (red dashed lines) in figure~\ref{fig:sgmfit}.  In particular at low redshifts, the differences are too large to be explained by the divergence of the Taylor expansion in terms of $k\mu$, which indicates some missing part in our understanding of FoG.

This missing part is the multi-streaming effect. As explained earlier,  $D^{\rm FoG}$ contains two parts, one induced by the bulk motion $D_{\rm bulk}$ and one induced by the multi-streaming effect or random motions inside halos $D_{\rm multi}$,
\bea
D^{\rm FoG}&=& D_{\rm bulk}D_{\rm multi} \,,\\
D_{\rm bulk}\simeq  e^{-k_z^2\sigma^2_{v,\rm bulk}/H^2} \,,
&\quad&
D_{\rm multi}\simeq e^{-k_z^2\sigma^2_{v,\rm multi}/H^2}\,.\no
\eea
Here the bulk velocity dispersion is by definition the measured velocity dispersion on regular grid points, the red dashed lines in figure~\ref{fig:sgmfit}. Therefore, the total FoG is indeed a Gaussian form,  with
\ba
\label{eqn:dispersion}
\sigma_v^2=\sigma^2_{v,\rm bulk}+\sigma^2_{v,\rm multi}\,.
\ea
Namely the velocity dispersion should be interpreted as the sum of the bulk flow and random motion. 
Figure~\ref{fig:sgmfit} shows that it is indeed the case. Namely, the inclusion of multi-streaming well explains the FoG measured by \cite{Zheng16a}. In particular, the agreement is excellent at all redshifts if the bulk flow velocity dispersion is given by the linear theory prediction (blue solid lines).  Furthermore, towards lower redshift, the contribution from multi-streaming becomes more and more significant, since $\sigma_{\rm multi}$ increases with decreasing redshift, while $\sigma_{\rm bulk}$ starts to decrease after $z\sim 0.5$ due to the slowed linear structure growth rate by cosmic acceleration. At $z=0$, we have $\sigma_{v, \rm multi}\simeq  260$ km/$s$,  very close to $\sigma_{v,\rm bulk}=306$ km$/s$. 

This excellent agreement means that we are able to physically understand $\sigma_v$, instead of treating it as a nuisance parameter in RSD. Since the structure growth rate $f\sigma_8$ causes enhancement of clustering while $\sigma_v$ damps the clustering strength, there exists severe degeneracy between $f\sigma_8$ and $\sigma_v$. This degeneracy can significantly degrade the RSD cosmological constraints. To break this degeneracy,  theoretical modelling of $\sigma_v$ is required. Figure~\ref{fig:sgmfit} shows that $\sigma^2_{v,\rm bulk}$ is well predicted by linear theory.  If we can further predict $\sigma^2_{v,\rm multi}$ from structure formation, $\sigma_v^2$ will no longer be a nuisance parameter. Instead it will be a source of useful cosmological information.  By definition, $\sigma^2_{v,\rm multi}$ arises from the non-linear evolution and we expect that the major contribution comes from virialized motion inside halos. Therefore,  in the next section we will adopt the halo model to gain understanding of the measured $D_{\rm multi}$.

\section{Explanation from halo model}
\label{sec:halomodel}
The halo model \cite{Cooray02} has been extended to redshift space by \cite{White01,Seljak01}. We follow these works to understand $D_{\rm multi}$ by the redshift space halo model. Assuming all dark matter particles reside in halos with different masses $M_i$ and positions $\bfx_i$, the dark matter density field $\rho(\bfx)$ could be written as
\beq
\label{eq:rho_x}
\rho(\bfx)=\sum_i f(\bfx-\bfx_i|M_i)\,,
\eeq
where $f(\bfx-\bfx_i|M_i)$ is the density profile of the $i$th halo. In Fourier space,
\bea
\label{eq:rho_k}
\rho(\bfk)&=&\int d^3\bfx\sum_i f(\bfx-\bfx_i|M_i)e^{i\bfk\cdot(\bfx-\bfx_i)}e^{i\bfk\cdot\bfx_i} \non \\
&=&\sum_i f(\bfk|M_i)e^{i\bfk\cdot\bfx_i}\,.
\eea
So the real space density power spectrum is calculated as 
\bea
\label{eq:pddreal}
P_{\delta\delta}(\bfk)&=&\left\langle\delta(\bfk)\delta^\ast(\bfk)\right\rangle=\frac{1}{\bar{\rho}^2}\left\langle\rho(\bfk)\rho^\ast(\bfk)\right\rangle \non \\
&=&\frac{1}{\bar{\rho}^2}\sum_{i,j}f(\bfk|M_i)f^\ast(\bfk|M_j)e^{i\bfk\cdot(\bfx_i-\bfx_j)} \non \\
&=&\frac{1}{\bar{\rho}^2}\sum_i |f(\bfk|M_i)|^2 + \frac{1}{\bar{\rho}^2}\sum_{i\neq j}f(\bfk|M_i)f^\ast(\bfk|M_j)e^{i\bfk\cdot(\bfx_i-\bfx_j)} \non \\
&=&\frac{1}{\bar{\rho}^2}\int |f(\bfk|M)|^2\frac{dn}{dM}dM \no \\
&&+\frac{1}{\bar{\rho}^2}\int P_{hh}(\bfk)f(\bfk|M_1)f^\ast(\bfk|M_2)\frac{dn}{dM_1}\frac{dn}{dM_2}dM_1dM_2\,, 
\eea
in which $P_{hh}$ is the real space halo-halo power spectrum.

In redshift space, on large scales, the peculiar velocity field distorts the halo-halo density power spectrum and replaces $P_{hh}$ with $P^s_{hh}$ in eq.~(\ref{eq:pddreal}). While on small scales the virial motions inside the halos distort the halo density profile and reduce the clustering power. The virial motion coexists with shell crossing and multi-streaming, so it is dominantly the source of the multi-streaming damping effect we try to measure in this paper. If the halos are assumed to be isotropic, virialized and isothermal with one-dimensional (1D) comoving velocity dispersion $\sigma_{v,\rm vir}$, then the random motions within the halos will effectively add a Gaussian damping window function $W_D$ to the halo density profile \cite{White01},
\beq
\label{eq:w_d}
W_D(k_z\sigma_{v,\rm vir})=e^{-k_z^2\sigma_{v,\rm vir}^2/2H^2(z)}\,,
\eeq
which leads the redshift space density power spectrum to be shown as
\bea
\label{eq:pddred}
P^s_{\delta\delta}&=&P^s_{\rm 1-halo}+P^s_{\rm 2-halo} \non\\
&=&\frac{1}{\bar{\rho}^2}\int |f(\bfk|M)|^2W^2_D(k_z\sigma_{v,\rm vir}|M)\frac{dn}{dM}dM \\
&&+\frac{1}{\bar{\rho}^2}\int P^s_{hh}(\bfk)f(\bfk|M_1)f^\ast(\bfk|M_2)W_D(k_z\sigma_{v,\rm vir,1}|M_1)W_D(k_z\sigma_{v,\rm vir,2}|M_2)\frac{dn}{dM_1}\frac{dn}{dM_2}dM_1dM_2 \,. \non
\eea

In this paper we do not attempt to calculate the above equation rigorously, but starting from it, we try to approximately discuss how well we could explain the measured $D_{\rm multi}$ in figure~\ref{fig:d_measure}. Therefore, we make an approximation that the multi-streaming damping could be characteristically estimated by the velocity dispersion inside halos  of a single characteristic mass $M_\ast$. Then the window function $W_D$ could be moved out of the integrals in eq.~(\ref{eq:pddred}) and we have
\bea
P^s_{\delta\delta}&=&W^2_D(k_z\sigma_{v,\rm vir}|M_\ast)\left[\frac{1}{\bar{\rho}^2}\int |f(\bfk|M)|^2\frac{dn}{dM}dM \right. \\
&&\left.+\frac{1}{\bar{\rho}^2}\int P^s_{hh}(\bfk)f(\bfk|M_1)f^\ast(\bfk|M_2)\frac{dn}{dM_1}\frac{dn}{dM_2}dM_1dM_2\right] \,. \non\\
&=&W^2_D(k_z\sigma_{v,\rm vir}|M_\ast)P^s_{\rm single} \non.
\eea

So we have
\beq
\label{eq:dapprox}
D_{\rm multi}\approx W^2_D(k_z\sigma_{v,\rm vir,\ast})=e^{-k_z^2\sigma_{v,\rm vir,\ast}^2/H^2(z)}\,.
\eeq
We adopt the Gaussian fitting formula for $D_{\rm multi}$ measurement in section~\ref{sec:measure}, then we have $\sigma_{v,\rm multi}=\sigma_{v,\rm vir,\ast}$. The 1D physical velocity dispersion $\sigma_{v,\rm phy,\ast}$ inside a halo with mass $M_\ast$ is
\beq
\label{eq:sigmav}
\sigma_{v,\rm phy,\ast}^2=a^2\sigma_{v,\rm vir,\ast}^2=a^2\sigma_{v,\rm multi}^2=\frac{GM_\ast}{2r_{\rm vir}}\,.
\eeq
We are more concerned about $\sigma_{v,\rm multi}^2/H^2$. Adopting the fitting formula provided by \cite{Bryan98,Cooray02}, it reads
\beq
\label{eq:fit_sgmv}
\frac{\sigma_{v,\rm multi}^2}{H^2}=\frac{\left[102.5\times f_{\sigma} \Delta^{1/6}_{\rm vir}(z)\left(\frac{H(z)}{H_0}\right)^{1/3}\left(\frac{M_\ast}{10^{13}M_\odot/h}\right)^{1/3}\right]^2}{a(z)^2H^2(z)}\,,
\eeq
where $\Delta_{\rm vir}=18\pi^2+82x-39x^2$ \cite{Bryan98}, with $x=\Omega_m(z)-1$ and $\Omega_m(z)=[\Omega_{m,0}(1+z^3)][H_0/H(z)]^2$, and $f_{\sigma}\sim 0.9$ is used to match the normalization from the simulations \cite{Bryan98}. 

\begin{figure}
\centering
\includegraphics[width=0.65\textwidth]{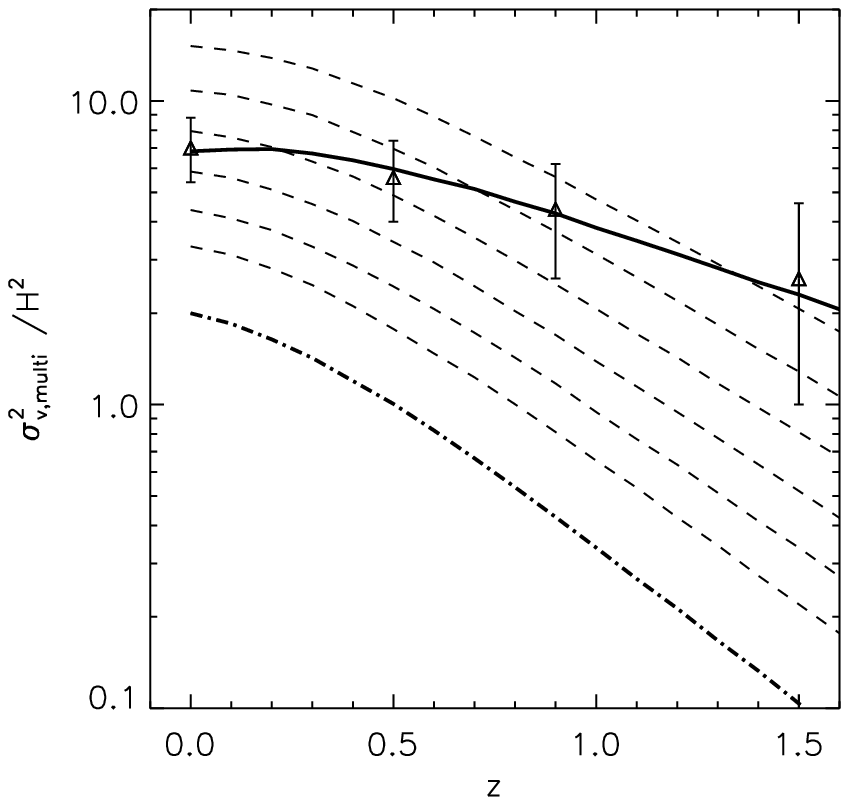}
\label{fig:sgmv2}
\caption{The data points with error bars are the fitted $\sigma_{v,\rm multi}^2/H^2$ of Gaussian function in the right panel of figure~\ref{fig:d_measure}. The solid line is for $\delta_\ast=1.25$ and $\gamma=0.65$. The dot-dashed line is for $\delta_\ast=1.686$ and $\gamma=1$. From top to bottom, the dashed lines are for $\delta_\ast=1.0,\ 1.1,\ 1.2,\ 1.3,\ 1.4,\ 1.5$ and $\gamma=1$.}
\end{figure}

$M_\ast$ depends on the halo mass function, which strongly relies on the r.m.s. density perturbation $\sigma(M,z)$,
\ba
\sigma^2(M,z)\equiv \int\frac{dk}{k}\frac{k^3P_{\rm lin}(k,z)}{2\pi^2}|W(kR)|^2\,, \no
\ea
where $W(x)= (3/x^3) [\sin(x)-x \cos(x)]$.
 A natural guess of $M_\ast$ is then 
\be
\label{eq:M_star}
\sigma(M_\ast,z)=\frac{\delta_\ast}{D(z)}\,,
\ee
 with $\delta_\ast\sim 1.686$ the spherical collapse threshold and $D(z)$ the linear density growth factor normalized as unity at $z=0$.  We use the code provided by \cite{Reed07} to calculate $\sigma(M,z)$, assuming the same cosmology of our simulations and $n_{\rm eff}$ dependence. Then we calculate $M_\ast$ by eq.~(\ref{eq:M_star}) and substitute it to eq.~(\ref{eq:fit_sgmv}) to calculate $\sigma_{v,\rm multi}^2/H^2$. The results with $\delta_*\in [1.0,1.686]$ are shown in figure~\ref{fig:sgmv2}. The choice of $\delta_\ast\sim 1.2$ shows reasonable agreement with the fitted multi-streaming $\sigma_{v,\rm multi}^2/H^2$, although the redshift dependence is too strong. 

The above recipe is too simplified to accurately describe multi-streaming, since in reality halos of various masses contribute to FoG and the relative contribution varies with redshift. Therefore it is not surprising that the redshift evolution of the fitted $\sigma_{v,\rm multi}^2/H^2$ deviates from the simple form above. This extra freedom in redshift evolution may be parametrized by 
\bea
\label{eq:sigmam}
\sigma(M_\ast,z)=\delta_\ast D(z)^{-\gamma}\,.
\eea
The choice of $\gamma=0.65$ and $\delta_\ast=1.25$ produces an excellent fit. More robust  investigation should follow eq.~(\ref{eq:pddred}), and will be presented elsewhere. 

\section{Conclusions and discussions}
\label{sec:con}




The multi-streaming effect measured in this paper is from dark matter particles. $\sigma_{v,\rm bulk}$ and $\sigma_{v,\rm multi}$ are interpreted as velocity dispersions on the grid and inside a grid cell.  So the measured damping effect depends on the grid size we choose: the smaller the grid size, the smaller the multi-streaming damping. The reason is due to the fact that the definition of bulk flow depends on the velocity assignment method and therefore depends on the grid size. 

In reality, RSD is measured from galaxies instead of dark matter particles. Therefore $\sigma_{v,\rm multi}$ should be interpreted as random motions of galaxies inside halos. It can differ from that of dark matter particles in the same halo and is expected to be smaller for either the reason of dynamical friction or being the central galaxies. This kind of velocity bias has been detected recently in local SDSS satellite galaxies by \cite{Guo15}.  This effect depends on the galaxy type and therefore brings extra uncertainty in its modelling. Recently, \cite{Okumura15b} considered the random motions of satellite galaxies in RSD modelling. 

A robust investigation associated with halo model can follow eq.~(\ref{eq:pddred}) and will be presented elsewhere. Also, eq.~(\ref{eq:sigmam}) will help us reduce the number of free parameters when we combine the data from different redshifts into one analysis.  Furthermore, the extra $k$ and/or $\mu$ dependence of $D_{\rm multi}$ away from a Gaussian function of $k\mu$ could help us disentangle $\sigma_{v,\rm bulk}$ and $\sigma_{v,\rm multi}$, if we extend the analysis to a large value of $k\mu$. This would be beneficial for either breaking the $f\sigma_8-\sigma_v$ degeneracy or constraining the galaxy evolution through constraining $\sigma_{v,\rm multi}$.

\section{Acknowledgments}

We thank Yong-Seon Song, Eric Linder and Uros Seljak for useful discussions. The work of running simulation was supported by the National Institute of Supercomputing and Network/Korea Institute of Science and Technology Information with supercomputing resources including technical support (KSC-2015-C1-017). Numerical calculations were performed by using a high performance computing cluster in the Korea Astronomy and Space Science Institute. PJZ thanks the support of the National
Science Foundation of China (11433001, 11320101002, 11621303),  National Basic Research Program of China (2015CB857001), Key Laboratory for Particle Physics, Astrophysics and Cosmology, Ministry of Education, and Shanghai Key Laboratory for Particle
Physics and Cosmology(SKLPPC)

\appendix
\section{Numerical artifacts treatment}
\label{appsec:numerical}
In this work we use the fast Fourier transform (FFT) on a regular grid to make Fourier transformation and calculate the power spectra. The first step of calculation is to sample the density and velocity fields on $512^3$ regular grid points. The discreteness of particles and the sampling procedure will introduce numerical artifacts to the measured power spectra.

\subsection{The density field sampling artifacts}
\label{appsubsec:d_sampling}
As derived in \cite{Jing05}, the measured density power spectrum $\hat{P}_{\delta\delta}(\bfk)$ is related to true density power spectrum $P_{\delta\delta}(\bfk)$ as
\bea
\label{eq:pk_sample}
\hat{P}_{\delta\delta}(\bfk)&=&\sum_{\bfn}|W(\bfk+2k_{\rm N}\bfn)|^2P_{\delta\delta}(\bfk+2k_{\rm N}\bfn) \non \\
&&+\frac{1}{n_{\rm P}}\sum_{\bfn}|W(\bfk+2k_{\rm N}\bfn)|^2\,.
\eea
Here $W(\bfk)$ is the Fourier transform of the mass assignment function $W(\bfr)$, $\bfn$ represents all three-dimensional integer vectors, $k_{\rm N}=\pi/H$ is the Nyquist frequency, with $H$ being the grid size, and $n_{\rm P}$ being particle number density. Since we use different power spectrum normalization from that of \cite{Jing05}, the shot-noise amplitude is determined by the particle number density $n_{\rm P}$ instead of the total particle number $N_{\rm P}$.

Eq.~(\ref{eq:pk_sample}) contains three numerical artifacts. The shot-noise $1/n_{\rm P}$ comes from the discreteness of the particles. The window function effect $W(\bfk)$ comes from the mass assignment procedure. The alias effect, that the measured power spectrum at $\bfk$ is affected by true power spectrum at $\bfk+2k_{\rm N}\bfn$, comes from the non-infinitesimal grid size we use for sampling.

The window function $W(\bfk)$ could be expressed as \cite{Hockney81,Jing05} 
\bea
W(\bfk)=\left[\frac{\sin(\pi k_1/2k_{\rm N})\sin(\pi k_2/2k_{\rm N})\sin(\pi k_3/2k_{\rm N})}{(\pi k_1/2k_{\rm N})(\pi k_2/2k_{\rm N})(\pi k_3/2k_{\rm N})}\right]^p\,,
\eea
where $k_{i=1,2,3}$ is the $i$th component of $\bfk$, and $p$ varies with different mass assignment functions: $p=1$ for the nearest grid point (NGP) function, $p=2$ for the clouds-in-cell (CIC) function, and $p=3$ for the triangular-shaped cloud (TSC) function.

The shot-noise term is different for different mass assignment method as well. According to \cite{Jing05}, for $k\leq k_{\rm N}$, we have 
\bea
\frac{1}{n_{\rm P}}\sum_{\bfn}W^2(\bfk+2k_{\rm N}\bfn)\equiv\frac{1}{n_{\rm P}}C_1(\bfk),\non \\
C_1(\bfk)\left\{
\begin{array}{ll}
=1& {\rm NGP\,,} \\
\approx 1-\frac{2}{3}\sin^2\left(\frac{\pi k}{2k_{\rm N}}\right) & {\rm CIC\,,} \\
\approx 1-\sin^2\left(\frac{\pi k}{2k_{\rm N}}\right) +\frac{2}{15}\sin^4\left(\frac{\pi k}{2k_{\rm N}}\right) & {\rm TSC\,.}
\end{array}\right.
\eea

When calculating $P_{\rm single}$, we separate the calculation into three power spectra, as shown in eq.~(\ref{eq:separate_p}). Assuming no velocity field sampling artifact, similar to \cite{Jing05}, we derive the relations between the measured $\hat{P}_1$/$\hat{P}_2$/$\hat{P}_3$ and the true $P_1$/$P_2$/$P_3$, as 
\bea
\hat{P}_1(\bfk)&=&\sum_{\bfn}|W(\bfk+2k_{\rm N}\bfn)|^2P_1(\bfk+2k_{\rm N}\bfn) \non \\
&&+\frac{1}{n_{\rm P}}\sum_{\bfn}|W(\bfk+2k_{\rm N}\bfn)|^2\,, \\
\hat{P}_2(\bfk)&=&\sum_{\bfn}W(\bfk+2k_{\rm N}\bfn)P_2(\bfk+2k_{\rm N}\bfn)\,, \\
\hat{P}_3(\bfk)&=&P_3(\bfk)\,.
\eea

Following these formulas, we correct the shot-noise and window function effect from the density related power spectra. For the alias effect, since it is most significant at $k\sim k_{N}=0.85$ for our simulation,  and our measurement is restricted to $k\lesssim0.2h$/Mpc, we consider it being negligible and skip its correction. If needed, we could follow the way of \cite{Jing05} to correct the alias effect.

\subsection{The velocity field sampling artifacts}
\label{appsubsec:v_sampling}

\begin{figure}
\centering
\includegraphics[width=0.65\textwidth]{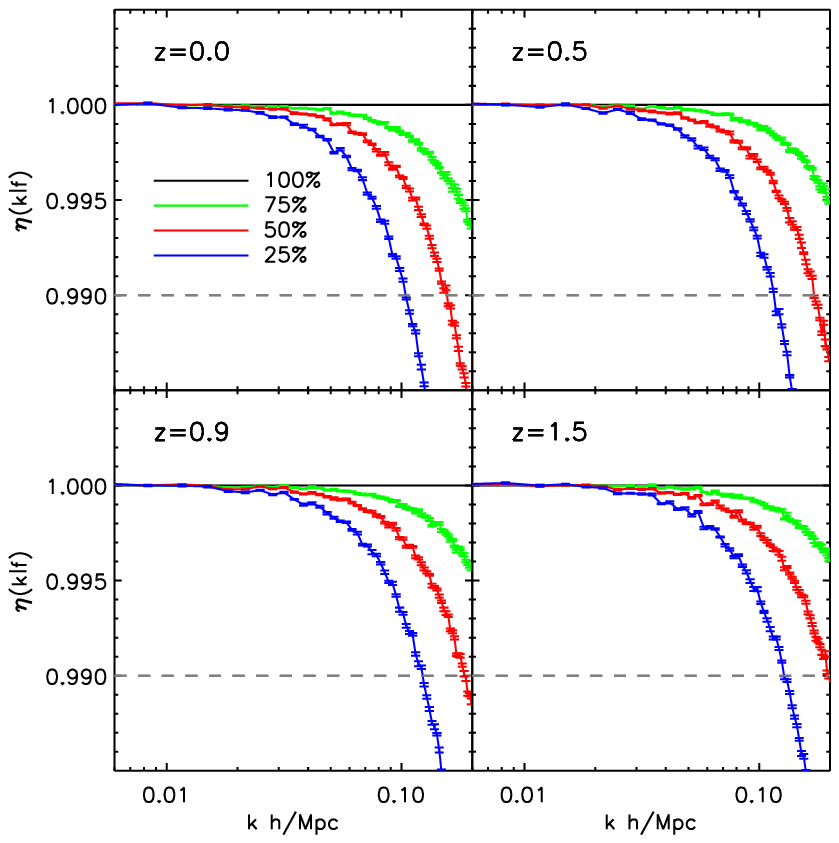}
\caption{The convergence test for the accuracy of our velocity field sampling by NP method.}
\label{fig:vtest}
\end{figure}

We use the volume-weighted velocity field in our analysis, which requires the fair sampling of the velocity field everywhere. Since we only have information of the velocity field where there is a particle or galaxy, the intrinsic clustering of inhomogeneous density field causes sampling artifacts of the measured velocity field, no matter what sampling methods we adopt \cite{Zheng13,Zheng14a,Zheng14b}.

The NP method we adopt indicates that we assign the velocity of the nearest particle/halo/galaxy to one grid point to this grid point position \cite{Zheng13}. Here we take the velocity power spectrum of an indication of the accuracy of the sampled velocity field. The systematic errors of the velocity power spectrum measurement by NP method, including window function effect and aliasing effect etc, have been tested and presented in the Appendix of \cite{Zheng13}. The sampling artifacts depend on the grid size and the particle number density \cite{Zheng13,Zheng14a,Zheng14b,Koda13}. For the grid size, the convergence tests shows that the velocity power spectrum is converged at $k\lesssim0.3h/$Mpc for  grid size equals to $4.7$Mpc/$h$. The smaller the grid size, the smaller the systematic error. Since the grid size of our choice in this paper is $3.69$Mpc/$h$, which is smaller than $4.7$Mpc/$h$, so the window function effect and aliasing effect, associated with the grid size will not affect our conclusion. For the sampling effect, originated from the inhomogeneous distribution of dark matter particles, we carry out a convergence test, the same as those in \cite{Zheng13,Zheng14a}, to check the accuracy of the velocity field sampled  by NP method in this work.

We randomly select a fraction $f$ of dark matter particles from the whole sample to construct a subsample. If there is no sampling artifact, its velocity power spectrum, denoted as $\hat{P}_{v}(k|f)$, will be identical to that of the whole sample, $\hat{P}_{v}(k|f=1)$. So we take the ratio
\bea
\eta(k|f)\equiv\frac{\hat{P}_{v}(k|f)}{\hat{P}_{v}(k|f=1)}
\eea
to measure the sampling artifact. We separately construct 10 subsamples with $f=75\%,\ 50\%,$ and  $25\%,$, and the measured $\eta(k|f)$ are shown in figure~\ref{fig:vtest}

We take $\eta(k|f=50\%)$ as an indicator of the sampling artifact, showing that our velocity power spectrum measurement is accurate within $1.0\%\sim1.5\%$ at $k\lesssim0.2h$/Mpc. At higher redshifts, the velocity measurement tends to be more accurate \cite{Zheng14a}.

The more accurate convergence test would be to directly calculate $\hat{P}_{\rm single}(k|f)$ from subsamples. Since the main purpose of this paper is to illustrate the importance and general properties of the multi-streaming effect. We assign the more rigorous test to the future study. Also the sampling artifact could be partially corrected following the method proposed in \cite{Zhang14,Zheng14a,Koda13}.

\bibliographystyle{JHEP}
\bibliography{mybib}
\end{document}